\documentclass[12pt]{article}   	
\usepackage{geometry}                		
\usepackage{graphicx}				
\usepackage{slashed}   
\usepackage{amssymb}
\usepackage{amsmath}
\usepackage{color}
\usepackage{framed}
\usepackage{cite}
\def\be{\begin{equation}}
\def\bea{\begin{eqnarray}}
\def\ee{\end{equation}}
\def\eea{\end{eqnarray}}
\def\hodge{\ast}

\begin{document}

\makeatletter
\def\blfootnote{\xdef\@thefnmark{}\@footnotetext}  
\makeatother

\begin{center}
{\LARGE Multi-center Superstrata}
\\
\vspace{18mm}
{\bf   Wukongjiaozi Tian}
\vspace{14mm}

Department of Physics,\\ University at Albany (SUNY),\\ Albany, NY 12222, USA\\ 

\vskip 10 mm

\blfootnote{jtian2@albany.edu}

\end{center}

\begin{abstract}

We construct a new class of superstrata, the regular supergravity solutions describing the microstates of D1-D5-P black holes. Our solutions are obtained by adding momentum charge to the D1-D5 geometries based on multiple concentric Kaluza-Klein monopoles.

\end{abstract}

\newpage

\section{Introduction}

The black hole information paradox is an important long-standing problem in theoretical physics \cite{Hawking:1976ra}. Although in the last two decades string theory has addressed some aspects of this question, starting with the counting of microscopic states of black holes \cite{Strominger:1996sh,Breckenridge:1996is}, the paradox has not been fully resolved. 
One of the approaches to the black hole information problem, known as the fuzzball program \cite{Lunin:2001jy,Mathur:2005zp,Bena:2007kg,Skenderis:2008qn,Chowdhury:2010ct}, suggests that microscopic states of black holes are represented by smooth and horizonless solutions of string theory. Some of such states can be seen already in the supergravity approximation, and all the two-charges solutions were constructed in  \cite{Lunin:2001jy,Lunin:2002iz}. However, to describe macroscopic black holes, one needs to find geometries carrying three charges\footnote{Alternatively, one may introduce non-extremality, but unfortunately non--supersymmetric regular geometries remain beyond reach with an exception of a very special solution found in \cite{Jejjala:2005yu}. See 
\cite{Roy:2016zzv} for the recent discussion of properties of this solution.}, and the last decade witnessed an impressive progress in 
this direction  \cite{Lunin:2004uu, Giusto:2004id,Giusto:2004kj,Bena:2005va,Berglund:2005vb,Bena:2006is,Lunin:2012gp}. \par

Recently a large class of D1-D5-P microstates geometries has been conjectured in \cite{Bena:2011uw,Bena:2014qxa}. These solutions depend on arbitrary functions of two variables, and they became known as superstrata. Some special superstrata have been constructed in \cite{Bena:2016agb,Bena:2015bea,Niehoff:2013kia,Giusto:2013bda,Shigemori:2013lta,Bena:2016ypk} by applying various generating techniques \cite{Bena:2013ora,Bena:2008wt,Ford:2006yb,Lunin:2004uu} to a two-charge round supertube \cite{Mateos:2001qs}. We will extend this construction by starting with a background produced by several concentric supertubes.  

This paper has the following organization. In section 2 we review the BPS equations governing  superstrata. 
In section 3 we will find the solutions describing the ``dressing'' of several concentric supertubes to a three--charge system, extending the results of \cite{Bena:2016agb,Bena:2015bea}. The solution will contain several free parameters, which will be determined in section 4 from the requirement of regularity. 

\section{Ansatz for the BPS ansatz solutions}

We work in the six dimensional truncation of type IIB supergravity on $\mathbb{M}^{4,1}\times \mathbb{S}^1\times \mathbb{T}^4$ and study the BPS solutions. The equations governing all supersymmetric geometries with trivial dependence on $T^4$ were written in \cite{Gutowski:2003rg,Giusto:2013rxa}, where it was shown that the metric must have the form
\begin{equation}
\label{eqn1}
ds_6^2=-\frac{2}{\sqrt{\mathcal{P}}}(dv+\beta)(du +\omega+\frac{\mathcal{F}}{2}(dv+\beta))+\sqrt{\mathcal{P}}ds_4^2(\mathcal{B}).
\end{equation}
Here $(\mathcal{P},\mathcal{F})$ are functions and $(\beta,\omega)$ are one--forms on the four--dimensional hyper--Kahler base $ds_4^2(\mathcal{B})$. As demonstrated in \cite{Niehoff:2013kia}, finding solutions with flat base reduces to a sequence of linear problems, and we will focus on such bases. 

The null coordinates in (\ref{eqn1}) are defined as
\begin{equation}
u=\frac{t-y}{\sqrt{2}},~~~v=\frac{t+v}{\sqrt{2}},\qquad y\sim y+2\pi R
\end{equation}
where $y$ is the parameter of $\mathbb{S}^1$. Using reparameterizations in (\ref{eqn1}), one can always choose $\mathcal{F}$ to be $v$-independent and vanishing at infinity, and later we will see that $\mathcal{F}$ is related to the momentum charge. 

We will choose the four dimensional metric of the base, $ds_4^2(\mathcal{B})$, to be in the Gibbons-Hawking (GH) form:
\begin{equation}
ds_{4}^2=V^{-1}(d\tau+A)^2+V(dy_1^2+dy_2^2+dy_3^2),
\end{equation}
where $\tau$ is the periodic parameter, function V and one-form $A$ satisfy the condition:
\begin{equation}
\hodge_3 d_3 \hodge_3 d_3V=0,~~~\hodge_3 d_3A=d_3V,
\end{equation}
with $d_3$ the differential and $\hodge_3$ the Hodge dual on $\mathbb{R}^3$. To fix the supersymmetric background we also need to find the one-form $\beta$ which is required
to have a self-dual field strength on the base:
\begin{equation}
\hodge_4\mathcal{D}\beta=\mathcal{D}\beta.
\end{equation}
Here the covariant derivative is defined as:
\begin{equation}
\mathcal{D}=d_4-\beta\wedge\frac{\partial}{\partial v},
\end{equation}
and later we will use dot to denote the derivative respect to $v$.

\par The field contents of this theory can be described in terms of three warp factors $Z_1$, $Z_2$ and $Z_4$ and three two-forms field strengths $\Theta_1$, $\Theta_2$ and $\Theta_4$. The BPS equations determine them are \cite{Bena:2011dd}:
\begin{equation}
\begin{aligned}
&\hodge_4\mathcal{D}\dot{Z}_1=\mathcal{D}\Theta_2,~\mathcal{D}\hodge_4\mathcal{D}Z_1=-\Theta_2\wedge \mathcal{D}\beta,~\Theta_2=\hodge_4\Theta_2,\\
&\hodge_4\mathcal{D}\dot{Z}_2=\mathcal{D}\Theta_1,~\mathcal{D}\hodge_4\mathcal{D}Z_2=-\Theta_1\wedge \mathcal{D}\beta,~\Theta_1=\hodge_4\Theta_1,\\
&\hodge_4\mathcal{D}\dot{Z}_4=\mathcal{D}\Theta_4,~\mathcal{D}\hodge_4\mathcal{D}Z_4=-\Theta_4\wedge \mathcal{D}\beta,~\Theta_4=\hodge_4\Theta_4.\\
\end{aligned}
\end{equation}
After solving these linear differential equations one should substitute them into another set of BPS equations to solve $\omega$ and $\mathcal{F}$ :
\begin{equation}
\begin{aligned}
&\mathcal{D}\omega+\hodge_4\mathcal{D}\omega+\mathcal{F}d\beta=Z_1\Theta_1+Z_2\Theta_2-2Z_4\Theta_4\\
&\hodge_{4}\mathcal{D}\hodge_{4}(\dot{\omega}+\frac{1}{2}\mathcal{D}\mathcal{F})=\ddot{\mathcal{P}}-(\dot{Z_1}\dot{Z_2}-(\dot{Z_4})^2)-\frac{1}{2}\hodge_{4}(\Theta_{1}\wedge\Theta_{2}-\Theta_{4}\wedge\Theta_{4})\\
&\mathcal{P}=Z_1Z_2-Z_4^2.
\end{aligned}
\end{equation}
We will summarize the result of \cite{Niehoff:2013kia} in the Appendix B.

\section{Local structure of the microstate geometries}
In this section we will construct three-charges BPS solutions describing fluctuating supertubes. A supertube is a supersymmetric configuration of D2-branes \cite{Mateos:2001qs}, which is characterized by a closed curve (profile) specifying the location of D2-branes. Upon duaization to the D1-D5 frame, the supertubes produce regular geometries constructed in \cite{Lunin:2001jy}. In the previous work a momentum charge has been added to a single circular supertube \cite{Bena:2016agb,Bena:2015bea,Niehoff:2013kia,Giusto:2013bda,Shigemori:2013lta,Bena:2016ypk}, and here we will extend these results to multiple concentric supertubes.
\subsection{Geometries with AdS$_3\times$S$^3$ asymptotics}
We start with a D1-D5 geometry sourced by concentric circular profiles. In this subsection we will add the momentum charge to solutions with AdS$_3\times$S$^3$ asymptotics, and this result will be extended to asymptotically-flat geometries in section 3.2.
For the four dimensional base we will still choose the flat metric:
\begin{equation}
ds_4^2=dr^2+r^2 d\theta^2+r^2\sin^2\theta d\phi^2+r^2\cos^2\theta d\psi^2, 
\end{equation}
but the one-form $\beta$ is sourced by many concentric circular supertubes so that the self-duality condition is satisfied trivially:
\begin{equation}
\begin{aligned}
&\beta=\frac{R}{\sqrt{2}}[(\frac{r^2+a^2-f_a}{2f_a})d\phi+(\frac{r^2-a^2-f_a}{2f_a})d\psi]+[a\leftrightarrow b]\\
&f_a=\sqrt{(r^2+a^2)^2-4r^2a^2\sin^2\theta},~~f_{b}=f_{a\leftrightarrow b},
\end{aligned}
\end{equation}
here we focus on the case of  two concentric circular supertubes with radii $a$ and $b$ to demonstrate the features of the solutions.  More general case can be found in the Appendix A. It is convenient to define some functions and self-dual two forms first:
\begin{equation}
\begin{aligned}
&\hat{v}=\frac{\sqrt{2}v}{R},~k_a=\sqrt{\frac{a^2+f_a-r^2}{a^2+f_a+r^2}},~k_b=k_{a\leftrightarrow b},\\
&\Omega^{(1)}=-\frac{r^3\sin2\theta}{f_a(a^2+r^2\cos2\theta+f_a)}(dr\wedge d\theta+r\sin\theta\cos\theta d\phi\wedge d\psi)+\big[a\leftrightarrow b],\\
&\Omega^{(2)}=\frac{-a^2+r^2+f_a}{2f_a}(\frac{dr\wedge d\psi}{r}+\tan\theta d\theta\wedge d\phi)\\
&~~~~~~~~~+\frac{a^2+r^2-f_a}{2f_a}(\frac{dr\wedge d\phi}{r}-\cot\theta d\theta\wedge d\psi)+\big[a\leftrightarrow b],
\end{aligned}
\end{equation} 
and note that
\begin{equation}
\Omega^{(1)}\wedge\Omega^{(2)}=0.
\end{equation}
The BPS equations can be solved by:
\begin{equation}
\label{eqnSuTb1}
\begin{aligned}
&Z_1=\frac{R^2}{Q}[\frac{a^2+c_1^2}{f_a}+\frac{b^2+c_2^2}{f_b}+c^2(k_ak_b)^{2n}\cos(2n(\hat{v}-2\psi))(\frac{1}{f_a}+\frac{1}{f_b})],\\
&Z_2=\frac{Q}{f_a}+\frac{Q}{f_b},~~Z_4=\sqrt{2}cR \cos(n(\hat{v}-2\psi)) (k_ak_b)^n(\frac{1}{f_a}+\frac{1}{f_b}),~~\Theta_1=0,\\
&\Theta_2=\frac{Rc^2}{\sqrt{2}Q}4n\,(k_ak_b)^{2n}[\sin(2n(\hat{v}-2\psi))\Omega^{(1)}-\cos(2n(\hat{v}-2\psi))\Omega^{(2)}],\\
&\Theta_4=2\,c\,n(k_ak_b)^n[\sin(2n(\hat{v}-2\psi))\Omega^{(1)}-\cos(2n(\hat{v}-2\psi))\Omega^{(2)}],\\
&\frac{\mathcal{F}}{2}=-\frac{2c^2}{r^2}(1-(k_ak_b)^{2n}),\\
&\omega=\beta_\phi d\phi-\beta_\psi d\psi-\frac{R\mathcal{F}}{\sqrt{2}}(\frac{1}{f_a}+\frac{1}{f_b})(r^2\sin^2\theta d\phi+r^2\cos^2\theta d\psi),
\end{aligned}
\end{equation}
where $c$ is an integration constant, and $c_1$ and $c_2$ are added in order to achieve the regularity later. The two terms in $Z_2$ corresponds two circular supertubes with D5--charge equal to $Q$.
The solutions have been written in a particular form in order to compare with the results in \cite{Giusto:2013bda}. Note that the supertubes are superposed in a very simple way, such simplicity persists for an arbitrary number of centers, and the relevant solution is written in the Appendix \ref{AppA}.
For coincident supertubes one recovers the solution constructed in \cite{Bena:2016agb}. Since 
$Z\sim 1/r^2$ at large values of $r$, solutions (\ref{eqnSuTb1}) and (\ref{SuTube2}) describe geometries with AdS$_3\times$S$^3$ asymptotics.

\bigskip
Let us demonstrate that solution (\ref{eqnSuTb1}) can be transformed to another frame, where it describes a supertube with charge oscillation added to a multi-center Gibbons-Hawking (GH) base. To see this, we
apply a spectral flow and spectral interchange in sequence:
\begin{equation}
\label{SpecFlow}
\begin{aligned}
&\hat{v}\rightarrow \hat{v}+\tau; ~~~~2\hat{v}\rightarrow -\tau,~\tau \rightarrow -2\hat{v}.\\
&\tau=\psi+\phi.
\end{aligned}
\end{equation}
The set of harmonic functions solving the first layer BPS equations 
becomes\footnote{We follow the conventions of \cite{Niehoff:2013kia}.}:
\begin{equation}
\begin{aligned}
&V=\frac{4}{f_a}+\frac{4}{f_b},~~K_1=\frac{2\sqrt{2}}{R}(\frac{Q}{f_a}+\frac{Q}{f_b}),~~
K_2=\frac{2\sqrt{2}R}{Q}(\frac{a^2+c_1^2}{f_a}+\frac{b^2+c_2^2}{f_b}),\\&K_3=\frac{\sqrt{2}R}{r^2},~~L_2=0,~~L_1=\frac{R^2c^2}{Q}\frac{(k_ak_b)^{2n}}{r^2}\cos(n(\tau-2\lambda)),~\lambda=\psi-\phi.
\end{aligned}
\end{equation}
The oscillating function $L_1$ can be understood as the oscillating field strength sourced by the charge density $\rho_1$ at the supertube location $r=0$ \cite{Bena:2010gg}:
\begin{equation}
L_1 = 4\pi \int d^3 y' \int d\tau' \hat{G}(\tau,\vec{y},\tau',\vec{y}')\rho_1 \delta^3(\vec{y}-\vec{y}')
\end{equation}
where $\hat{G}$ is the Green function on the the base, and for a general GH geometry such function was constructed in  \cite{Page:1979ga}. The answer for our two-center base is
\begin{equation}
\hat{G}(\tau,\vec{y},\tau',\vec{y}')=\frac{1}{4\pi^2r^2}\frac{(k_ak_b)^{-2}}{(k_ak_b)^{-2}-\cos(\frac{1}{2}(\tau-\tau')-(\lambda-\lambda'))}\,.
\end{equation}
The added charge densities can be identified by matching the Fourier's modes:
\begin{equation}
\rho_1=\frac{R^2c^2}{4Q}\cos n(\tau-2\lambda).
\end{equation}
We infer that in this frame the solution describes a singly-wound supertube added charge density $\rho_1$ in a two-center GH base. We conclude this subsection by observing that solution (\ref{eqnSuTb1}) can also be generated by procedures presented in \cite{Bena:2016agb}, but one needs to introduce an additional spectral flow like (\ref{SpecFlow}).

 \subsection{Asymptotically--flat geometries}
To extend the solution (\ref{eqnSuTb1}) to the asymptotically--flat region, we need to make a modification:
\begin{equation}
Z_1\rightarrow 1+Z_1,~Z_2\rightarrow 1+Z_2,~\omega\rightarrow  \omega +\delta\omega,
\end{equation}
where $\delta\omega$ has to satisfy \cite{Giusto:2013bda}:
\begin{equation}
\mathcal{D}\delta \omega+\hodge_4\mathcal{D}\delta\omega=\Theta_2,~~\hodge_4\mathcal{D}\hodge_4\delta\omega=\dot{Z}_1.
\end{equation}
Substituting into the general  $Z_1$ and $\Theta_2$ with $N$ centers from the Appendix A, one finds
\begin{equation}
\label{eqn21}
\begin{aligned}
&\delta\omega=\frac{Rc^2}{\sqrt{2}Q}k_\alpha^{2n}[\sin(2n\eta)\omega^{(0)}-\cos(2n\eta)d\psi]\\
&\omega^{(0)}=\frac{dr}{r}-\tan\theta d\theta,~~k_\alpha\equiv\prod_{i=1}^{N}k_{a_i},~~\eta\equiv \hat{v}-N\psi
\end{aligned}
\end{equation}
This expression has a singularity at $r=0$ which can be eliminated by adding homogenous solutions or suitable Fourier modes of the warp factors. We will cure all the singularities in next section. For future reference, here we present the final regular asymptotically--flat geometry:
\begin{equation}
\begin{aligned}
&ds_4^2=dr^2+r^2d\theta^2+r^2\sin^2\theta d\phi^2+r^2\cos^2\theta d\psi^2,~~\beta=\sum_{i=1}^{N}\beta_i,\\
&Z_1=1+\frac{R^2}{Q}[\sum_{i=1}^{N}\frac{a_i^2+2c^2}{f_{a_1}}+c^2(1-\frac{\alpha^2}{Q})\frac{k_\alpha^{2n}\cos(2n\eta)}{f_{\alpha}}],\\
&Z_2=1+\frac{Q}{f_{\alpha}},~~Z_4=\sqrt{2}cR\frac{k_\alpha^n\cos(n\eta)}{f_{\alpha}},~~\frac{\mathcal{F}}{2}=-\frac{2c^2}{r^2}(1-k_\alpha^{2n}),\\
&\omega=\beta_\phi d\phi-\beta_\psi d\psi-\frac{R\mathcal{F}}{\sqrt{2}}\frac{1}{f_{\alpha}}(r^2\sin^2\theta d\phi+r^2\cos^2\theta d\psi)\\
&-\frac{Rc^2}{\sqrt{2}Q}k_\alpha^{2n}[\sin(2n\eta)(\alpha^2\sum_{i=1}^N\omega^{(1)}_{a_i}-\omega^{(0)})+\cos(2n\eta) (d\psi-
\sum_{i=1}^N\frac{\alpha^2\beta_i}{a_i^2})],\\
&~\omega^{(1)}_{a_i}=\frac{1}{f_{a_i}}(\frac{dr}{r}-\frac{a^2_i\sin2\theta d\theta}{r^2+a_i^2\cos2\theta+f_{a_i}}),~~\frac{1}{f_\alpha}\equiv\sum\frac{1}{f_{a_i}},~~\frac{1}{\alpha^2}\equiv\sum_{i=1}^{N}\frac{1}{a_{i}^2}.
\end{aligned}
\end{equation}

\section{Regular microstate geometries}
In this section we will study the potential singularities at the center of $\mathbb{R}^4$ and the positions of the supertubes, and we will make a minimal change of the warp factors to eliminate the singularities  encountered in last section.

We begin with analyzing  the regularity of the metric at the center of $\mathbb{R}^4$, i.e. at $r=0$. The only potential singular term is the function $\mathcal{F}$. Expand $\mathcal{F}$ around $r=0$ we can get:
\begin{equation}
\mathcal{F}\sim -4c^2n\frac{1}{\alpha^2}+O(r^2),
\end{equation}
which is regular. 

Next we focus on the singular points at the supertube locations, $f_{a_i}=0$. The potential singularities are in the coefficient of $(d\tau+A)^2$ in the six dimensional metric. The condition to cancel it is the requirement \cite{Bena:2015bea}:
\begin{equation}
\lim_{f_{a_i}\rightarrow 0}f_{a_i}[-\frac{2}{\sqrt{\mathcal{P}}}\beta_0(\mu+\frac{\mathcal{F}}{2}\beta_0)+\frac{\sqrt{\mathcal{P}}}{V}]=0,
\end{equation}
where $\beta_0=(\beta_\phi+\beta_\psi)/2$ and $\mu=(\omega_\phi+\omega_\psi)/2$. This condition will fix the values of $c_i$ in $Z_1$. One can find
\begin{equation}
c_i^2=2c^2.
\end{equation}

Finally we eliminate singularities in (\ref{eqn21}) by introducing a minimal change of the sources 
\begin{equation}
Z_1\rightarrow Z_1+\delta_1 Z_1,~~\delta_1 Z_1=\frac{R^2c_\delta}{Q} \frac{k_\alpha^{2n}\cos(2n\eta)}{f_{\alpha}}.
\end{equation}
This new sources will induce a modification to the one-form $\omega$ which will can make  (\ref{eqn21})  regular. Solving the second set of BPS equations, we find
\begin{equation}
\begin{aligned}
&\delta_1 \omega_{r,\theta}=\frac{Rc_\delta}{\sqrt{2}}k_\alpha^{2n}\sin(2n\eta)\sum_{i=1}^N\omega^{(1)}_{a_i},\\
&\delta_1 \omega_{\phi,\psi}=-\frac{Rc_\delta}{\sqrt{2}}k_\alpha^{2n}\cos(2n\eta)(\sum_{i=1}^N\frac{\beta_i}{a_i^2}),\\
&\delta_1\mathcal{F}=0,~~
\omega^{(1)}_{a_i}=\frac{1}{f_{a_i}}(\frac{dr}{r}-\frac{a^2_i\sin2\theta d\theta}{r^2+a_i^2\cos2\theta+f_{a_i}}).
\end{aligned}
\end{equation}
Regularity of the one--form $\delta\omega+\delta_1\omega$  at $r=0$ fixes the value of $c_\delta$:
\begin{equation}
c_\delta=-\frac{c^2\alpha^2}{Q}.
\end{equation}
This elimination of the singularities is typical for superstrata, and it is known as coiffuring \cite{Bena:2014rea}.
\section{Discussion}
This work generalizes the superstrata discovered in \cite{Bena:2016agb} by adding momentum to 
concentric D1--D5 profiles. 
In some frame our solution describes a supertube with charge oscillation added in a multi-center Gibbons-Hawking base, and as argued in \cite{Bena:2008wt}, such multi-centered solutions describe unbound states of the D1--D5--P system. Surprisingly, the full geometry is obtained by taking a simple superposition of the harmonic functions describing individual centers, and it would be interesting to demonstrate that a similar superposition principle holds for all unbound states.

\section{Acknowledgements}
I would like to thank my advisor Oleg Lunin for suggesting this project and his guidance throughout this work. This work was supported by NSF grant PHY-1316184 and by the Physics Department at the University at Albany.

\appendix

\section{Solution with arbitrary number of supertubes}
\label{AppA}
In this appendix we present the solution with many concentric supertubes.
Following the pattern encountered in (\ref{eqnSuTb1}) it is easy to guess the geometry with $N$ concentric supertubes,
\begin{equation}
\label{SuTube2}
\begin{aligned}
&\beta=\sum_{i=1}^{N}\beta_i,\\
&Z_1=\frac{R^2}{Q}[\sum_{i=1}^{N}\frac{a_i^2+c_i^2}{f_{a_1}}+c^2(\prod_{i=1}^{N}k_{a_i})^{2n}\cos(2n(\hat{v}-N\psi))\sum\frac{1}{f_{a_i}}],\\
&Z_2=Q\sum_{i=1}^{N}\frac{1}{f_{a_i}},~~~Z_4=\sqrt{2}cR\cos(n(\hat{v}-N\psi))(\prod_{i=1}^N k_{a_i})^n\sum_{i=1}^{N}\frac{1}{f_{a_i}},\\
&\frac{\mathcal{F}}{2}=-\frac{2c^2}{r^2}(1-(\prod_{i=1}^N k_{a_i})^{2n}),\\
&\omega=\beta_\phi d\phi-\beta_\psi d\psi-\frac{R\mathcal{F}}{\sqrt{2}}\sum_{i=1}^N\frac{1}{f_{a_i}}(r^2\sin^2\theta d\phi+r^2\cos^2\theta d\psi),\\
\end{aligned}
\end{equation}
and check that it solves all the BPS equations.
In the language of two-charge geometry, the solution has $N$ concentric circular profiles with radii $a_i$.

\section{Solution generating technique}
\label{AppB}
In this appendix we review the work of \cite{Niehoff:2013kia} and show some details in our construction.
The 1-form $\beta$ appearing in (\ref{eqn1}) can be separated as:
\begin{equation}
\beta=\frac{K_3}{V}(d\tau+A)+\xi
\end{equation}
with $K_3$ is a harmonic function on $\mathbb{R}^3$ and its relation with 1-form $\xi$ is:
\begin{equation}
\hodge_3 d_3\xi=-d_3K_3
\end{equation}
The first layer of the solutions can be built as
\begin{equation}
\begin{aligned}
&\Theta_{2}=\mathcal{D}(\frac{K_{2}}{V})\wedge(d\tau+A)+\hodge_{4}[\mathcal{D}(\frac{K_{2}}{V})\wedge(d\tau+A)]\\
&Z_{1}=L_{1}+\frac{K_{2}K_{3}}{V}
\end{aligned}
\end{equation}
where $K_{2}$ and $L_{1}$ are generalized harmonic functions satisfy:
\begin{equation}
\hodge_{4}\mathcal{D}\hodge_{4}K_{2}=\hodge_{4}\mathcal{D}\hodge_{4}L_{1}=0,
\end{equation}
with a constraint:
\begin{equation}
\partial_{\tau}K_{2}+\partial_{v}L_{1}=0.
\end{equation}
Similarly we have:
\begin{equation}
\begin{aligned}
&\Theta_{1}=\mathcal{D}(\frac{K_{1}}{V})\wedge(d\tau+A)+\hodge_{4}[\mathcal{D}(\frac{K_{1}}{V})\wedge(d\tau+A)],\\
&Z_{2}=L_{2}+\frac{K_{1}K_{3}}{V},\\
&\hodge_{4}\mathcal{D}\hodge_{4}K_{1}=\hodge_{4}\mathcal{D}\hodge_{4}L_{2}=0,\\
&\partial_{\tau}K_{1}+\partial_{v}L_{2}=0.
\end{aligned}
\end{equation}
and

\begin{equation}
\begin{aligned}
&\Theta_{4}=\mathcal{D}(\frac{K_{4}}{V})\wedge(d\tau+A)+\hodge_{4}[\mathcal{D}(\frac{K_{4}}{V})\wedge(d\tau+A)],\\
&Z_{4}=L_{4}+\frac{K_{4}K_{3}}{V},\\
&\hodge_{4}\mathcal{D}\hodge_{4}K_{4}=\star_{4}\mathcal{D}\hodge_{4}L_{4}=0,\\
&\partial_{\tau}K_{4}+\partial_{v}L_{4}=0.
\end{aligned}
\end{equation}
To crucial step solving the first layer BPS equation is to solve the generalized harmonic function given the $\beta$. For example let us solve $K_4$.
The equation about $K_4$ is
\begin{equation}
\hodge_4\mathcal{D}\hodge_4\mathcal{D} K_4=0
\end{equation}
where $\mathcal{D}=d-\beta\wedge\frac{\partial}{\partial v}$.
The attempt solution is $K_4=e^{i(mv+k\phi+n\psi)}f(r,\theta)$. 
\begin{equation}
\mathcal{D}K_4=dK_4-im\beta K_4,~~\hodge\mathcal{D}K_4=\hodge dK_4-im K_4 \hodge\beta 
\end{equation}
\begin{equation}
\mathcal{D}\hodge\mathcal{D}K_4=d\hodge dK_4-im dK_4\wedge\hodge\beta-im K_4 d\hodge \beta-\beta\wedge(im\hodge dK_4+m^2 K_4\hodge\beta)
\end{equation}
Consider a special $\beta$:
\begin{equation}
\beta=\beta_\phi(r,\theta) d\phi+\beta_\psi(r,\theta) d\psi,~~\hodge\beta=\sqrt{g}(\beta_\phi\epsilon^\phi_{r\theta\psi}dr\wedge d\theta\wedge d\psi+\beta_\psi\epsilon^\psi_{r\theta\phi}dr\wedge d\theta\wedge d\phi)
\end{equation}
which implies $d\hodge \beta=0$. With the equality $\beta\wedge\hodge dK_4=dK_4\wedge\hodge\beta$ in the end
we get:
\begin{equation}
\hodge d\hodge dK_4+2m(k g^{\phi\phi}\beta_\phi+ng^{\psi\psi}\beta_\psi)K_4-m^2(g^{\phi\phi}\beta_\phi^2+g^{\psi\psi}\beta_\psi^2)K_4=0
\end{equation}
The family of solution vanishing at infinity is:
\begin{equation}
K_4=\frac{e^{i(k\phi-mv+n\psi)}}{r^2}(\frac{\sin\theta}{r})^{k}(\frac{\cos\theta}{r})^{n}(h_{ra}h_{rb})^m
\end{equation}
where 
\begin{equation}
h_{ra}=\sqrt{\frac{r^2}{(-a^2+f_a+r^2)}\frac{r^2}{(a^2+f_a+r^2)}}
\end{equation}
To eliminate the singularities let us rewrite $\cos\theta$:
\begin{equation}
\cos\theta=\sqrt{\frac{(-a^2+f_a+r^2)(a^2+f_a-r^2)}{4r^2a^2}}=\sqrt{\frac{(-b^2+f_b+r^2)(b^2+f_b-r^2)}{4r^2b^2}}
\end{equation}
and set $m=2n$ and $k=0$:
\begin{equation}
K_4=\frac{e^{i(-nv+2 n\psi)}}{r^2}k_a^nk_b^n
\end{equation}
where
\begin{equation}
k_a=\sqrt{\frac{a^2+f_a-r^2}{a^2+f_a+r^2}}.
\end{equation}

\end{document}